\documentclass[reprint, onecolumn, nofootinbib, amsmath, amssymb, aps, amsfonts, amsthm, bm]{revtex4-2}

\usepackage{graphicx}
\usepackage{dcolumn}
\usepackage{bm}

\begin{document}

\title{Fluid Dynamics Beyond the Continuum - A Physical Perspective on Large Eddy Simulation}

\author{Max Okraschevski}
\email{E-Mail: max.okraschevski@kit.edu}
\author{Sven Hoffmann}
\author{Katharina Stichling}
\author{Rainer Koch}
\author{Hans-Joerg Bauer}
\affiliation{Institute of Thermal Turbomachinery, Karlsruhe Institute of Technology, Kaiserstraße 12, 76131 Karlsruhe, Germany}

\date{\today}

\begin{abstract}
In this work, we will present a physically consistent theory to derive the governing equations of the Large Eddy Simulation (LES) framework based on first principles rather than the motivation to conduct computationally affordable simulations of turbulent flows. Therefore, we assume that a coarse-grained fluid element, subsequently called super fluid element, can be locally defined comprising a large number of smaller elementary fluid elements. Then, similar to non-equilibrium molecular dynamics (NEMD), in which the transport equations of an elementary fluid element can be consistently reconstructed from the local, collective dynamics of molecules, the transport equations of a super fluid element can be derived from the local, collective dynamics of elementary fluid elements. Interestingly, we find: (a) Favre filtering is a physical consistency condition, (b) why Boussinesq's hypothesis in conjunction with eddy viscosity models is commonly employed in LES and (c) that the LES framework might be more than a numerical turbulence model for computational fluid dynamics (CFD).
\end{abstract}

\maketitle

\emph{Introduction} - It is well known from non-equilibrium molecular dynamics (NEMD) that the fundamental transport equations of fluid dynamics can be consistently reconstructed from the collective dynamics of individual molecules by means of an appropriate averaging process. The first theory providing such a link between the microscopic and macroscopic world was developed by Irving and Kirkwood \cite{Irving_1950}. Using a statistical mechanics approach, which requires the a priori knowledge of the multivariate probability density function (pdf) of the molecule collective describing its phase space distribution, they were able to define an appropriate averaging process by moments of the pdf. Although their theory is brilliant from a theoretical point of view, for practical considerations the determination of the pdf is hardly possible as it originally relies on ensemble statistics. Solutions were developed to circumvent the ensemble averages, and other minor issues, yet Hardy was the first one presenting an alternative theory avoiding all these issues at the same time \cite{Hardy_1982}.  According to Hardy theory, the averaging process is alternatively defined by introduction of a local, symmetrical, positive and monotonously decaying kernel function $W_h(\mathbf{r}) \in \mathbb{R}$ with compact support $supp\{W_h(\mathbf{r})\} \subset \mathbb{R}^3$. The quantity $\mathbf{r} \in \mathbb{R}^3$ denotes a distance vector and the index $h \in \mathbb{R}$ is a parameter that determines the local extent of the compact support domain. For the sake of simplicity, we will assume for the rest of this work that the shape of the support is represented either by a sphere or a cube, and that $h=const$ throughout the fluid domain of interest. Further claiming that the kernel is normalized

\begin{equation}
    \int\displaylimits_{supp\{W_h(\mathbf{r})\}} W_h(\mathbf{r})~\mathrm{d}\mathbf{r} = 1~,
    \label{eq:KernelNormalization}
\end{equation}

and that the individual molecules $j$ at position $\mathbf{y}_j(t) \in \mathbb{R}^3$ evolve in (non-negative) time $t \in \mathbb{R}^+_0$ with constant mass $M_j \in \mathbb{R}$ , Hardy defined the density at the fixed position $\mathbf{x} \in \mathbb{R}^3$ according to a locally weighted average

\begin{equation}
    \rho(\mathbf{x}, t) := \sum_{j=1}^{N} W_h(\mathbf{x}-\mathbf{y}_j(t)) M_j~.
    \label{eq:DensityFluidElement}
\end{equation}

\begin{figure}[t]
\centering
\includegraphics[trim= 0cm 0cm 0cm 0cm, width=7in]{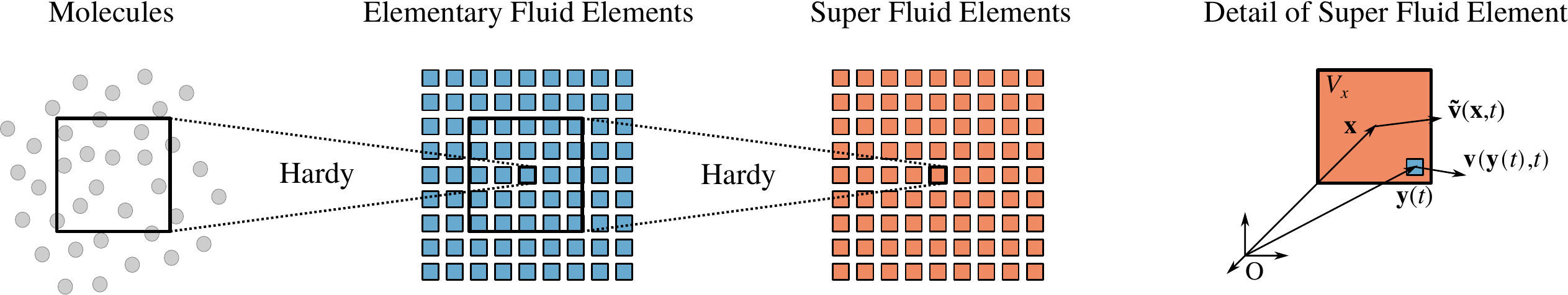}
\caption{Schematic of fluid dynamics beyond the continuum. By consecutive application of Hardy theory to a sufficiently large number of molecules and, further, elementary fluid elements, one first obtains the local dynamics of elementary fluid elements (blue) and finally the dynamics of coarse-grained super fluid elements (orange).}
\label{fig:Superfluidelement}
\end{figure}

In Eq. (\ref{eq:DensityFluidElement}) the integer $N$ represents the number of molecules inside the local support domain, which obviously has to be sufficiently large such that the value converges towards the density of the elementary fluid element. Then, by consistent evaluation of the Eulerian temporal derivative of Eq. (\ref{eq:DensityFluidElement}), namely $\partial_t \rho(\mathbf{x}, t)$, the continuity equation of fluid dynamics can be reconstructed providing a natural microscopic definition of the macroscopic velocity $\mathbf{v}(\mathbf{x}, t) \in \mathbb{R}^3$ and the corresponding mass flux $\dot{\mathbf{m}} (\mathbf{x}, t) =  \rho(\mathbf{x}, t) \mathbf{v}(\mathbf{x}, t) \in \mathbb{R}^3$. With the latter, a subsequent evaluation of the Eulerian temporal derivative $\partial_t \dot{\mathbf{m}} (\mathbf{x}, t)$ leads to a reproduction of the fluid dynamic momentum transport equation with a microscopic definition of the stress tensor $\boldsymbol{\sigma}(\mathbf{x}, t) \in \mathbb{R}^{3\times3}$. As demonstrated in Hardy's original work, the procedure can be repeated in order to construct the macroscopic transport equation for energy \cite{Hardy_1982} and it can be applied to obtain species transport equations. Vividly speaking, we can summarize that the Hardy theory is the mathematical framework for the concept which is usually depicted in fluid mechanics textbooks to illustrate an elementary fluid element comprising a large number of molecules (FIG. \ref{fig:Superfluidelement}). Hence, it transfers smaller particles (molecules) into coarse-grained particles (elementary fluid elements) satisfying axiomatic conservation properties scale-independently. \\
Subsequently, we present the outcome of Hardy theory consecutively applied on an even coarser length scale to elementary fluid elements instead of molecules. This idea, as well as the notation utilized hereinafter, is illustrated in FIG. \ref{fig:Superfluidelement}. Accordingly, we have to introduce the concept of a super fluid element, which comprises an infinitely large number of elementary fluid elements. Interestingly, during the construction of the transport equations for the super fluid element, we rediscover the governing equations of the Large Eddy Simulation (LES) framework for compressible flows and well-known LES concepts \cite{Sagaut_2006, Garnier_2009, Pope_2011, Ferziger_2020}. This is demonstrated in the following, while restricting ourselves to the transport of mass and momentum only.

Although the interpretation of LES as a coarse-graining method is established in the literature, e.g. \cite{Geurts_2008, Parish_2017, Eyink_2018}, the coarse-graining was, to the authors' knowledge, not utilized as a mathematical starting point before to rigorously derive the governing equations of LES. Instead, even in the most popular textbooks dealing with LES, e.g. \cite{Sagaut_2006, Garnier_2009, Pope_2011, Ferziger_2020}, a filter is applied to the Navier-Stokes equation in order to separate scales of turbulent flows. Traditionally, LES is thus motivated based on Eulerian fields. The novelty and essence of our approach is that it motivates LES using Lagrangian particles, namely elementary fluid elements, on the basis of an established NEMD method.  This proves to be advantageous in that it can initially be considered independent of turbulence and eventually reveals that the super fluid element counterpart of the kinetic stress tensor $\boldsymbol{\tau}_{kin}$, emerging from Hardy theory, is inherently linked to turbulence and equivalent to the LES subgrid scale tensor $\boldsymbol{\tau}_{SGS}$. \\

\emph{Mathematical Notation} - In order to facilitate the comprehension of our following derivation, we shortly explain our mathematical notation which, especially in terms of Lebesgue integration, follows the framework presented by DiBenedetto \cite{DiBenedetto_2011}. As our derivation starts at the continuum level, we assume that our fluid domain of interest $\Omega \subset \mathbb{R}^3$ is a connected space and can be decomposed into an infinite number of elementary fluid elements as depicted in FIG. \ref{fig:Superfluidelement}. Whereas we denote a fixed Eulerian position as $\mathbf{x} \in \Omega$, the position of an elementary fluid element for a certain fixed time $t$ is given by $\mathbf{y}(t) \in \Omega$. If $t$ is instead interpreted as a variable parameter, which is subsequently necessary for the computation of temporal derivatives, then $\mathbf{y}(t)$ also represents the parametrization of individual elementary fluid element trajectories.

Further, we can assign the differential volume elements $\mathrm{d}\mathbf{y}(t)$ and differential mass elements $\mathrm{d}M(\mathbf{y}(t))$ to each individual elementary fluid element. Both are differential 3-forms and can be different for each elementary fluid element at $\mathbf{y}(t)$, which is possible if we interpret them in terms of Lebesgue measures \cite{DiBenedetto_2011}. This also motivates us to note both differential forms with explicit $\mathbf{y}(t)$ dependence. Then, in the framework of Lebesgue integration, the two differential forms are coupled by the local fluid density $\rho$, namely $\mathrm{d}M(\mathbf{y}(t)) = \rho(\mathbf{y}(t), t)\mathrm{d}\mathbf{y}(t)$, which represents a Radon–Nikodym derivative \cite{DiBenedetto_2011}.

With the former definitions, we can now introduce the locally weighted spatial average, which we will use subsequently as the coarse-graining method for the definition of super fluid element quantities. Mathematically, in the Lebesgue framework, we understand it as continuum analogy of the original discrete version proposed by Hardy. With $f: \Omega \times \mathbb{R}^+_0 \to \mathbb{R}$ denoting a scalar field, and $V_x := supp\{W_h(\mathbf{x} - \mathbf{y}(t))\}$ denoting a super fluid element centered at the fixed position $\mathbf{x}$ and defined by its local kernel support, the locally weighted spatial average reads

\begin{equation}
    \overline{f} (\mathbf{x}, t) = \int\displaylimits_{V_x} f(\mathbf{y}(t), t) W_h(\mathbf{x} - \mathbf{y}(t))~\mathrm{d}\mathbf{y}(t)~,
    \label{eq:LocalAverage1}
\end{equation}
where we denote the average or, alternatively, the super fluid element quantity with an overline. The locally weighted average is thus equivalent to the usual LES filtering \cite{Sagaut_2006, Garnier_2009, Pope_2011, Ferziger_2020}. We will use the overline notation in the sense of Eq. (\ref{eq:LocalAverage1}) only in our derivation for abbreviation purposes or to demonstrate the equivalency of our results to established LES equations. Instead, as every quantity of interest is attached to mass elements $\mathrm{d}M(\mathbf{y}(t))$, we actively work with the following locally weighted average, which always includes the density

\begin{equation}
    \overline{\rho f} (\mathbf{x}, t) = \int\displaylimits_{V_x} \rho f(\mathbf{y}(t), t) W_h(\mathbf{x} - \mathbf{y}(t))~\mathrm{d}\mathbf{y}(t) = \int\displaylimits_{V_x} f(\mathbf{y}(t), t) W_h(\mathbf{x} - \mathbf{y}(t))~\mathrm{d}M(\mathbf{y}(t))~.
    \label{eq:LocalAverage2}
\end{equation}
The locally weighted average in Eq. (\ref{eq:LocalAverage2}) can straightforwardly be extended to vector and tensor fields by component wise application. \\

\emph{Mass Transport} - We start with the derivation of the continuity equation for the super fluid element. Therefore, we consider Eq. (\ref{eq:DensityFluidElement}) in the limit $N \to \infty$. Consequently, as the elementary fluid elements span a continuum, the summation can be replaced by an integration. This means we can discard the index $j$ of the Lagrangian elementary fluid element positions $\mathbf{y}_j(t) \to \mathbf{y}(t)$ and the individual masses $M_j$ can be replaced by their differential counterparts $\mathrm{d}M(\mathbf{y}(t))$. The density of the super fluid element $V_x$, in the sense of Eq. (\ref{eq:LocalAverage2}) with $f (\mathbf{x}, t) = 1$, then reads

\begin{equation}
    \overline{\rho} (\mathbf{x}, t) = \int\displaylimits_{V_x} W_h(\mathbf{x} - \mathbf{y}(t))~\mathrm{d}M(\mathbf{y}(t))~.
    \label{eq:DensitySuperFluid1}
\end{equation}

In order to obtain the continuity equation of the super fluid element, the Eulerian temporal derivative $\partial_t \overline{\rho}$ has to be evaluated. Recalling that the  domain of integration $V_x$ is static and fixed in space, as well as that $\mathrm{d}M(\mathbf{y}(t))$ is unchanged along the Lagrangian flow path of an elementary fluid element as a direct consequence of continuity, we can interchange integration and temporal derivative yielding

\begin{equation}
    \partial_t \overline{\rho} (\mathbf{x}, t) = \int\displaylimits_{V_x} \partial_t W_h(\mathbf{x} - \mathbf{y}(t))~\mathrm{d}M(\mathbf{y}(t))~.
    \label{eq:TempDensitySuperFluid1}
\end{equation}

The term $\partial_t W_h(\mathbf{x} - \mathbf{y}(t))$ in Eq. (\ref{eq:TempDensitySuperFluid1}) can then be recast by application of the chain rule leading to

\begin{equation}
        \partial_t \overline{\rho} (\mathbf{x}, t) = \int\displaylimits_{V_x} \nabla_{\mathbf{x} - \mathbf{y}(t)} W_h(\mathbf{x} - \mathbf{y}(t)) \cdot [-\dot{\mathbf{y}}(t)]~\mathrm{d}M(\mathbf{y}(t))~.
    \label{eq:TempDensitySuperFluid2}
\end{equation}

In Eq. (\ref{eq:TempDensitySuperFluid2}) the index of nabla indicates the variables on which the operator is applied, the symbol "$\cdot$" denotes an inner product and $\partial_t \mathbf{y}(t) = \dot{\mathbf{y}}(t)$, as the position of a single elementary fluid element depends only on time. Further, we can kinematically identify the velocity of the elementary fluid element $\mathbf{v}(\mathbf{y}(t), t)=\dot{\mathbf{y}}(t)$ \footnote{ As described in the section \emph{Mathematical Notation}, we here interpret $\mathbf{y}(t)$ as the parametrization of the elementary fluid element trajectory since a temporal derivative is involved. As $\mathbf{v}$ represents the velocity field of the fluid flow, the vector field has to explicitly depend on $\mathbf{y}(t)$ in order to match the velocity of a specific, individual elementary fluid element. } and that $\nabla_\mathbf{x} W_h(\mathbf{x} - \mathbf{y}(t)) = \nabla_{\mathbf{x} - \mathbf{y}(t)} W_h(\mathbf{x} - \mathbf{y}(t))$ by using the chain rule again, such that

\begin{equation}
        \partial_t \overline{\rho} (\mathbf{x}, t) = -\int\displaylimits_{V_x} \nabla_{\mathbf{x}} W_h(\mathbf{x} - \mathbf{y}(t)) \cdot \mathbf{v}(\mathbf{y}(t), t) ~\mathrm{d}M(\mathbf{y}(t))~.
    \label{eq:TempDensitySuperFluid3}
\end{equation}

Since $\mathbf{v}(\mathbf{y}(t), t)$ is not depending on $\mathbf{x}$, the expression $\nabla_{\mathbf{x}} W_h(\mathbf{x} - \mathbf{y}(t)) \cdot \mathbf{v}(\mathbf{y}(t), t) = \nabla_{\mathbf{x}} \cdot [\mathbf{v}(\mathbf{y}(t), t)   W_h(\mathbf{x} - \mathbf{y}(t))]$ holds and, as the integration is applied with respect to elementary fluid element positions $\mathbf{y}(t)$, integration and $\nabla_{\mathbf{x}}$ can be interchanged again. Additionally using $\mathrm{d}M(\mathbf{y}(t)) = \rho(\mathbf{y}(t), t)\mathrm{d}\mathbf{y}(t)$, we finally arrive at the filtered continuity equation of LES, which describes mass conservation of a super fluid element

\begin{equation}
        \partial_t \overline{\rho} (\mathbf{x}, t) = - \nabla_{\mathbf{x}} \cdot \int\displaylimits_{V_x}  \rho\mathbf{v}(\mathbf{y}(t), t) W_h(\mathbf{x} - \mathbf{y}(t))   ~\mathrm{d}\mathbf{y}(t) = - \nabla_{\mathbf{x}} \cdot \int\displaylimits_{V_x}  \dot{\mathbf{m}}(\mathbf{y}(t), t) W_h(\mathbf{x} - \mathbf{y}(t))   ~\mathrm{d}\mathbf{y}(t) = - \nabla_{\mathbf{x}} \cdot ~ \overline{\dot{\mathbf{m}}}(\mathbf{x}, t)~.
    \label{eq:ContinuitySuperFluid}
\end{equation}

Based on Eq. (\ref{eq:ContinuitySuperFluid}), the question arises with which representative velocity $\tilde{\mathbf{v}}(\mathbf{x}, t)$ \footnote{We denote the velocity with a tilde as, generally, this velocity will not coincide with the locally weighted averaged (filtered) velocity of elementary fluid elements.} a super fluid element is moving? For the reason of physical consistency, we require that this representative velocity is mass conserving. This implies that the chosen velocity should preserve the mathematical structure of the traditional continuity equation. Then the only option for the definition is $\overline{\dot{\mathbf{m}}}(\mathbf{x}, t) = \overline{\rho} \tilde{\mathbf{v}}(\mathbf{x}, t)$, which naturally leads to the Favre filtered velocity \cite{Ferziger_2020}

\begin{equation}
    \tilde{\mathbf{v}}(\mathbf{x}, t) := \frac{\overline{\dot{\mathbf{m}}}(\mathbf{x}, t)}{\overline{\rho}(\mathbf{x}, t)} = \frac{\overline{\rho\mathbf{v}}(\mathbf{x}, t)}{\overline{\rho}(\mathbf{x}, t)}~.
    \label{eq:FavreVelocity}
\end{equation}

The insight that Favre filtering according to Eq. (\ref{eq:FavreVelocity}) is inherently mass conserving is not new and was previously stated by Bilger \cite{Bilger_1975}. However, our definition emerges from a physical reasoning, finally inducing the velocity of super fluid elements rather than a practical need to avoid correlation terms \cite{Bilger_1975}. \\

\emph{Momentum Transport} - With the former results, we can now compute the Eulerian temporal derivative of the mass flux  $\overline{\dot{\mathbf{m}}}(\mathbf{x}, t)$ from Eq. (\ref{eq:ContinuitySuperFluid}) in order to obtain the momentum transport equation of the super fluid element. By the same arguments as before, we absorb the elementary fluid density into the differential form and then interchange the temporal derivative and the integral

\begin{equation}
    \partial_t \overline{\dot{\mathbf{m}}}(\mathbf{x}, t) = \partial_t \overline{\rho\mathbf{v}}(\mathbf{x}, t) = \int\displaylimits_{V_x} \partial_t [  \mathbf{v}(\mathbf{y}(t), t) W_h(\mathbf{x} - \mathbf{y}(t))] ~\mathrm{d}M(\mathbf{y}(t))~.
    \label{eq:TempMomentumSuperFluid1}
\end{equation}

The integrand of Eq. (\ref{eq:TempMomentumSuperFluid1}) can be reformulated by the chain rule

\begin{equation}
    \partial_t [  \mathbf{v}(\mathbf{y}(t), t) W_h(\mathbf{x} - \mathbf{y}(t))] = \mathbf{v}(\mathbf{y}(t), t)\partial_t W_h(\mathbf{x} - \mathbf{y}(t)) + W_h(\mathbf{x} - \mathbf{y}(t))\partial_t \mathbf{v}(\mathbf{y}(t), t)~.
    \label{eq:TempMomentumSuperFluid1Integrand}
\end{equation}

For the derivation of the mass transport equation for the super fluid element, Eq. (\ref{eq:ContinuitySuperFluid}), we already rearranged the temporal derivative $\partial_t W_h(\mathbf{x} - \mathbf{y}(t))$ in Eq. (\ref{eq:TempMomentumSuperFluid1Integrand}). We proceed here likewise. For the second temporal derivative term in Eq. (\ref{eq:TempMomentumSuperFluid1Integrand}) we realize that the velocity of a specific elementary fluid element is a function of time only, namely $\mathbf{v}(t) = \mathbf{v}(\mathbf{y}(t), t)$. Consequently, we can kinematically identify the acceleration of an elementary fluid element $\mathbf{a}(\mathbf{y}(t), t) = \partial_t \mathbf{v}(\mathbf{y}(t), t)$ and then, in combination with Eq. (\ref{eq:TempMomentumSuperFluid1Integrand}) and Eq. (\ref{eq:TempMomentumSuperFluid1}), we find

\begin{equation}
    \partial_t \overline{\rho\mathbf{v}}(\mathbf{x}, t) = - \int\displaylimits_{V_x}   \mathbf{v}(\mathbf{y}(t), t) [\mathbf{v}(\mathbf{y}(t), t)  \cdot \nabla_{\mathbf{x}} W_h(\mathbf{x} - \mathbf{y}(t))]  ~\mathrm{d}M(\mathbf{y}(t)) + \int\displaylimits_{V_x} \mathbf{a}(\mathbf{y}(t), t)  W_h(\mathbf{x} - \mathbf{y}(t)) ~ \mathrm{d}M(\mathbf{y}(t))~.
    \label{eq:TempMomentumSuperFluid2}
\end{equation}

Using tensor calculus, we can now rewrite the first integrand on the right hand side (rhs) of Eq. (\ref{eq:TempMomentumSuperFluid2}) by introducing the tensor field $\mathbf{v}\mathbf{v}^T \in \mathbb{R}^{3 \times 3}$ with the superscript "$T$", indicating the transposed vector field. This tensor field operates then on $\nabla_{\mathbf{x}} W_h(\mathbf{x} - \mathbf{y}(t))$, giving for the first integrand of Eq. (\ref{eq:TempMomentumSuperFluid2}) 

\begin{equation}
    - \int\displaylimits_{V_x}   \mathbf{v}(\mathbf{y}(t), t) [\mathbf{v}(\mathbf{y}(t), t)  \cdot \nabla_{\mathbf{x}} W_h(\mathbf{x} - \mathbf{y}(t))]  ~\mathrm{d}M(\mathbf{y}(t)) = - \int\displaylimits_{V_x} \mathbf{v}\mathbf{v}^T (\mathbf{y}(t), t) \nabla_{\mathbf{x}} W_h(\mathbf{x} - \mathbf{y}(t))  ~\mathrm{d}M(\mathbf{y}(t)) ~.
    \label{eq:TempMomentumSuperFluid2Integrand}
\end{equation}

Since the tensor valued quantity $\mathbf{v}\mathbf{v}^T (\mathbf{y}(t), t)$ for a specific elementary fluid element does not depend on $\mathbf{x}$ and as the integration is applied with respect to elementary fluid element positions $\mathbf{y}(t)$, we can interchange the integration and the $\nabla_{\mathbf{x}}$ operator. Thus, the latter turns into a divergence operator $div_{\mathbf{x}}$ acting on the resulting tensor field. We then obtain, together with Eq. (\ref{eq:TempMomentumSuperFluid2Integrand}) and Eq. (\ref{eq:TempMomentumSuperFluid2})

\begin{equation}
    \partial_t \overline{\rho\mathbf{v}}(\mathbf{x}, t) = - div_{\mathbf{x}} \int\displaylimits_{V_x}  \mathbf{v}\mathbf{v}^T (\mathbf{y}(t), t)  W_h(\mathbf{x} - \mathbf{y}(t))  ~\mathrm{d}M(\mathbf{y}(t)) + \int\displaylimits_{V_x} \mathbf{a}(\mathbf{y}(t), t)  W_h(\mathbf{x} - \mathbf{y}(t)) ~ \mathrm{d}M(\mathbf{y}(t))~.
    \label{eq:TempMomentumSuperFluid3}
\end{equation}

With the notation of the locally weighted average from Eq. (\ref{eq:LocalAverage2}), we finally arrive at the momentum transport equations for the super fluid element, which coincide with the LES filtered momentum equations \cite{Garnier_2009, Ferziger_2020}

\begin{equation}
    \partial_t \overline{\rho\mathbf{v}}(\mathbf{x}, t) = - div_{\mathbf{x}} \left[ \overline{\rho\mathbf{v}\mathbf{v}^T} \right] (\mathbf{x}, t) + \overline{\rho \mathbf{a}} (\mathbf{x}, t)~,
    \label{eq:MomentumSuperFluid}
\end{equation}

where the individual accelerations $\mathbf{a}(\mathbf{y}(t), t)$ of the elementary fluid elements are given by the rhs of their corresponding momentum transport equation, e.g. Navier-Stokes for a Newtonian fluid \cite{Landau_1991}. But generally, this approach is independent of the rheological behaviour of the fluid of interest. \\

\emph{Kinetic Stress} - Usually, Eq. (\ref{eq:MomentumSuperFluid}) is claimed to be a conditional equation for the super fluid element velocity $\tilde{\mathbf{v}}(\mathbf{x}, t)$. Hence, we end up with the problem of correlation terms well-known from turbulence theory, either from Reynolds Averaged Navier Stokes (RANS) or LES \cite{Ferziger_2020}. This closure problem is traditionally solved by reformulation of the correlation terms in Eq. (\ref{eq:MomentumSuperFluid}). We will subsequently only discuss the reformulation of $\overline{\rho\mathbf{v}}$ and $\overline{\rho\mathbf{v}\mathbf{v}^T}$ and refer to well-known textbooks for the remaining term $\overline{\rho \mathbf{a}}$, e.g. \cite{Garnier_2009}, as it does not strengthen our physical interpretation. From the derivation of the continuity equation for the super fluid element we already have $\overline{\dot{\mathbf{m}}} = \overline{\rho\mathbf{v}} = \overline{\rho} \tilde{\mathbf{v}}$, which eliminates the first correlation issue. For the second correlation, we use standard LES reasoning, adding a zero by

\begin{equation}
    \overline{\rho\mathbf{v}\mathbf{v}^T} = \overline{\rho\mathbf{v}\mathbf{v}^T} -  \overline{\rho} \tilde{\mathbf{v}} \tilde{\mathbf{v}}^T + \overline{\rho} \tilde{\mathbf{v}} \tilde{\mathbf{v}}^T ~.
    \label{eq:ZeroLES}
\end{equation}

The term $\boldsymbol{\tau}_{SGS} := \overline{\rho\mathbf{v}\mathbf{v}^T} -  \overline{\rho} \tilde{\mathbf{v}} \tilde{\mathbf{v}}^T$ in Eq. (\ref{eq:ZeroLES}) is called subgrid scale stress (SGS) tensor in LES \cite{Ferziger_2020, Garnier_2009}. Combining the latter with Eq. (\ref{eq:ZeroLES}) and putting the result into Eq. (\ref{eq:MomentumSuperFluid}), the final form of the momentum transport equations of compressible LES can be recovered, based on first principles only. Obviously, it is just another form of the momentum transport equations for the super fluid element, which now gives $\tilde{\mathbf{v}}$ as required

\begin{equation}
    \partial_t [\overline{\rho} \tilde{\mathbf{v}}](\mathbf{x}, t) + div_{\mathbf{x}} \left[ \overline{\rho} \tilde{\mathbf{v}} \tilde{\mathbf{v}}^T \right] (\mathbf{x}, t) = - div_{\mathbf{x}} \left[ \boldsymbol{\tau}_{SGS} \right](\mathbf{x}, t) + \overline{\rho \mathbf{a}} (\mathbf{x}, t)~.
    \label{eq:MomentumSuperFluidLES}
\end{equation}

Instead of classically attempting to find a specific model for $\boldsymbol{\tau}_{SGS}$ \cite{Ferziger_2020}, we rather demonstrate that the subgrid stress tensor $\boldsymbol{\tau}_{SGS}$ is a super fluid element counterpart of the kinetic stress tensor $\boldsymbol{\tau}_{kin}$ emerging from Hardy theory.
Therefore, analogously to Hardy theory, we introduce the peculiar velocities $\mathbf{w}(\mathbf{x}, \mathbf{y}(t), t)$, which may change inside the super fluid element centered at $\mathbf{x}$, as it is indicated by the argument $\mathbf{y}(t)$. It is defined as the relative velocity between different elementary fluid elements and the corresponding super fluid element velocity for a given domain $V_x$ (FIG. (\ref{fig:Superfluidelement})), namely

\begin{equation}
    \mathbf{v}(\mathbf{y}(t), t) = \tilde{\mathbf{v}}(\mathbf{x}, t) + \mathbf{w}(\mathbf{x}, \mathbf{y}(t), t)~.
    \label{eq:PeculiarVelocity}
\end{equation}

Following a similar argument like Vreman \cite{Vreman_1994}, we can now isolate the definition of $\boldsymbol{\tau}_{SGS}$ and combine it with the velocity decomposition of Eq. (\ref{eq:PeculiarVelocity}) to obtain

\begin{eqnarray}
    \boldsymbol{\tau}_{SGS} (\mathbf{x}, t) &=& \overline{\rho\mathbf{v}\mathbf{v}^T}(\mathbf{x}, t) -  \overline{\rho} \tilde{\mathbf{v}} \tilde{\mathbf{v}}^T(\mathbf{x}, t) = \int\displaylimits_{V_x} \mathbf{v}(\mathbf{y}(t), t)\mathbf{v}^T (\mathbf{y}(t), t)  W_h(\mathbf{x} - \mathbf{y}(t))  ~\mathrm{d}M(\mathbf{y}(t)) - \overline{\rho} \tilde{\mathbf{v}} \tilde{\mathbf{v}}^T \nonumber \\
    &=& \int\displaylimits_{V_x} \left[ \tilde{\mathbf{v}} \tilde{\mathbf{v}}^T(\mathbf{x}, t) +  \tilde{\mathbf{v}}(\mathbf{x}, t) \mathbf{w}^T(\mathbf{x}, \mathbf{y}(t), t) + \mathbf{w}(\mathbf{x}, \mathbf{y}(t), t)\tilde{\mathbf{v}}^T(\mathbf{x}, t) + \mathbf{w}\mathbf{w}^T(\mathbf{x}, \mathbf{y}(t), t) \right] W_h(\mathbf{x} - \mathbf{y}(t))~ \mathrm{d}M(\mathbf{y}(t)) \nonumber \\
    &-& \overline{\rho} \tilde{\mathbf{v}} \tilde{\mathbf{v}}^T(\mathbf{x}, t)
    \label{eq:SGSDerivationKin1}~.
\end{eqnarray}

As the super fluid element velocity $\tilde{\mathbf{v}}(\mathbf{x}, t)$ does not depend on $\mathbf{y}(t)$, Eq. (\ref{eq:SGSDerivationKin1}) is equivalent to

\begin{eqnarray}
    \boldsymbol{\tau}_{SGS} (\mathbf{x}, t) &=&  \tilde{\mathbf{v}} \tilde{\mathbf{v}}^T(\mathbf{x}, t) \int\displaylimits_{V_x} W_h(\mathbf{x} - \mathbf{y}(t))~ \mathrm{d}M(\mathbf{y}(t)) -  \overline{\rho} \tilde{\mathbf{v}} \tilde{\mathbf{v}}^T(\mathbf{x}, t) \nonumber \\
    &+& \tilde{\mathbf{v}}(\mathbf{x}, t) \int\displaylimits_{V_x} \mathbf{w}^T(\mathbf{x}, \mathbf{y}(t), t)  W_h(\mathbf{x} - \mathbf{y}(t))~ \mathrm{d}M(\mathbf{y}(t)) + \left[ \int\displaylimits_{V_x} \mathbf{w}(\mathbf{x}, \mathbf{y}(t), t)  W_h(\mathbf{x} - \mathbf{y}(t)) ~ \mathrm{d}M(\mathbf{y}(t)) \right]\tilde{\mathbf{v}}^T(\mathbf{x}, t) \nonumber \\
    &+& \int\displaylimits_{V_x} \mathbf{w}\mathbf{w}^T(\mathbf{x}, \mathbf{y}(t), t)  W_h(\mathbf{x} - \mathbf{y}(t)) ~ \mathrm{d}M(\mathbf{y}(t)) ~,
    \label{eq:SGSDerivationKin2}
\end{eqnarray}

where we rearranged the last term in the third line of Eq. (\ref{eq:SGSDerivationKin1}). With Eq. (\ref{eq:DensitySuperFluid1}), the integrand in the first line of Eq. (\ref{eq:SGSDerivationKin2}) coincides with the density $\overline{\rho}$ of the super fluid element. Thus, the first line of Eq. (\ref{eq:SGSDerivationKin2}) is zero. Further, it can be verified that the second line of Eq. (\ref{eq:SGSDerivationKin2}) is zero as well due to vanishing peculiar momenta $\tilde{\mathbf{v}} \overline{\rho \mathbf{w}^T}$ and $\overline{\rho \mathbf{w}} \tilde{\mathbf{v}}^T$. This results from the locally weighted average of the peculiar velocity $\mathbf{w}(\mathbf{x}, \mathbf{y}(t), t)$ in  Eq. (\ref{eq:PeculiarVelocity}), for which

\begin{eqnarray}
    \int\displaylimits_{V_x} \mathbf{w}(\mathbf{x}, \mathbf{y}(t), t)  W_h(\mathbf{x} - \mathbf{y}(t)) ~ \mathrm{d}M(\mathbf{y}(t)) &=& \int\displaylimits_{V_x} \mathbf{v}(\mathbf{y}(t), t)  W_h(\mathbf{x} - \mathbf{y}(t)) ~ \mathrm{d}M(\mathbf{y}(t))  - \tilde{\mathbf{v}}(\mathbf{x}, t) \int\displaylimits_{V_x} W_h(\mathbf{x} - \mathbf{y}(t)) ~ \mathrm{d}M(\mathbf{y}(t)) \nonumber \\
    &=& \overline{\rho \mathbf{v}} (\mathbf{x}, t) - \overline{\rho}\tilde{\mathbf{v}}(\mathbf{x}, t) = 0~,
    \label{eq:SGSDerivationKin3}
\end{eqnarray}

as a consequence of continuity for the super fluid element (Eq. (\ref{eq:FavreVelocity})). Additionally, the transpose of Eq. (\ref{eq:SGSDerivationKin3}) holds too. Hence, the first and second line of Eq. (\ref{eq:SGSDerivationKin2}) vanish and we finally find the prove that $\boldsymbol{\tau}_{SGS}$ is a super fluid element counterpart of the kinetic stress tensor $\boldsymbol{\tau}_{kin}$ emerging from Hardy theory. It is the local covariance tensor of the velocity field

\begin{equation}
    \boldsymbol{\tau}_{SGS} (\mathbf{x}, t) = \int\displaylimits_{V_x} \mathbf{w}\mathbf{w}^T(\mathbf{x}, \mathbf{y}(t), t)  W_h(\mathbf{x} - \mathbf{y}(t)) ~ \mathrm{d}M(\mathbf{y}(t)) =: \boldsymbol{\tau}_{kin} (\mathbf{x}, t) ~.
    \label{eq:KineticStressSuperFluid}
\end{equation}

With this result, it finally becomes reasonable why Boussinesq's hypothesis and corresponding eddy viscosity models are commonly employed in the LES community. It is due to the fact that the original kinetic stress from Hardy theory and our super fluid element counterpart in Eq. (\ref{eq:KineticStressSuperFluid}) share the same mathematical structure. As the original Hardy stress tensor converges towards the macroscopic stress tensor in the thermodynamic limit \cite{Admal_2010}, which also incorporates the viscous shear stress tensor for Newtonian fluid flows, why should it not be possible to follow a similar approach for a super fluid element? Although such an approach seems promising at first glance due to the given reasoning, it is nowadays well-known that classical eddy viscosity models suffer from several issues, e.g. \cite{Schmitt_2007, Moser_2021}. This is a direct consequence of the unjustified modeling assumption that the turbulent mixing scale and the scale of the coarse-grained field variation are well separated. Nevertheless, we can imagine that the insight of the equivalence of $\boldsymbol{\tau}_{SGS} = \boldsymbol{\tau}_{kin}$ could be valuable for LES subgrid scale modeling of non-Newtonian fluids.

Furthermore, we want to complete this section on the kinetic stress tensor $\boldsymbol{\tau}_{kin}$ in Eq. (\ref{eq:KineticStressSuperFluid}) discussing its connection to turbulence. This question is inevitable at this point as so far the whole derivation was only motivated by coarse-graining and the determination of the dynamics of a super fluid element. However, the procedure results in the governing equations of LES which are utilized to model turbulent flows. The LES momentum transport equations mainly differ from the classical momentum transport equations in respect to the additional stress term emerging in Eq. (\ref{eq:MomentumSuperFluidLES}). Hence, it is crucial to understand under which conditions the kinetic stress tensor $\boldsymbol{\tau}_{kin}$ in Eq. (\ref{eq:KineticStressSuperFluid}) becomes significant. As $\boldsymbol{\tau}_{kin}$ is a quadratic form in the peculiar velocities $\mathbf{w}(\mathbf{x}, \mathbf{y}(t), t)$, the influence will be large either if the size of super fluid element is such that on the chosen length scale nonlinearities do matter or the fluid flow on the contiuum scale is already strongly nonlinear and hence turbulent \footnote{Using generic velocity fields, it can be  verified  that for a constant velocity field $\boldsymbol{\tau}_{kin}= \mathbf{0}$ and for a linear velocity field $\boldsymbol{\tau}_{kin}= \mathbf{const}$. In both cases the resulting force $div [ \boldsymbol{\tau}_{kin} ] = 0$.}. This implies that turbulence in our theory seems to be inherently linked to the kinetic stress tensor $\boldsymbol{\tau}_{kin}$ in Eq. (\ref{eq:KineticStressSuperFluid}) and the question arises if it is possible to introduce a quantitative measure for turbulence based on $\boldsymbol{\tau}_{kin}$ in Eq. (\ref{eq:KineticStressSuperFluid}). It may be worthwhile to have a deeper look.\\

\emph{Implications} - As modern numerical and experimental fluid dynamicists usually employ the coarse-grained super fluid element perspective, namely focusing on discretization volumes or measurement volumes, we believe that the implications of our super fluid element theory could have a far reaching impact. This becomes evident when considering the following examples:

\begin{itemize}

    \item Although Reynolds' 1895 paper is accepted as the foundation of the RANS framework, nowadays defined on the basis of ensemble averages \cite{Pope_2011}, Reynolds himself, contrary to the modern interpretation, suggested local volume averages, symbolically denoted by a "$\sum$" operator \cite{Reynolds_1895}.  Hence, we are convinced that our coarse-graining strategy matches Reynolds original idea, albeit Reynolds was not able to rigorously derive his equations of "mean motion" with the aid of Hardy theory, which was developed roughly a century after Reynolds' 1895 paper. As demonstrated in this work, the outcome of this derivation interestingly leads to a contradiction. Against the common expectation, the governing equations of LES and not modern RANS are obtained. Consequently, one may ask whether RANS and LES are physically different after all, because both might just be different mathematical ways to derive the transport equations of a super fluid element. On the one hand, we see strong evidence in regard to the modeling of turbulence in CFD codes. There, employing either of the methods is solely different in terms of the additional stress model, which simultaneously serves as an explanation for the success of hybrid RANS-LES approaches \cite{Heinz_2020}. On the other hand, the significance of non-locality concepts in RANS theory in order to correctly capture turbulence characteristics is indisputable, e.g. two-point correlations to define characteristic length scales \cite{Pope_2011}, structure functions, which constitute the success of Kolmogorov theory \cite{Pope_2011}, and the non-locality of the Reynolds stress tensor itself \cite{Hamba_2005}. This non-local character of turbulence is inherently included in Eq. (\ref{eq:KineticStressSuperFluid}). Hence, the idea to develop a unified, consistent mathematical framework for RANS and LES seems promising. Moreover, we believe such an attempt is likely to be successful, since Admal and Tadmor \cite{Admal_2010} were also able to demonstrate in the NEMD community that the original theory of Irving and Kirkwood, in analogy to RANS employing ensemble averages, and Hardy, in analogy to LES employing locally weighted spatial averages, can be consistently unified. This would imply that not only the LES subgrid stress tensor $\boldsymbol{\tau}_{SGS}$  but also the Reynolds stress $\boldsymbol{\tau}_{Rey}$ would turn out to be identical to the kinetic stress tensor $\boldsymbol{\tau}_{kin}$ in terms of a well-defined mathematical theory.
    
    \item Even for Particle Image Velocimetry (PIV) there seems to be a relation to our super fluid element theory. The goal of the PIV technique is to experimentally determine the velocity field of a flow by means of tracer particles, whose positions are cross-correlated within so called interrogation windows \cite{Westerweel_2013}. In order to evaluate the uncertainty of the local velocity vector, a commonly utilized method is the sub-window technique \cite{Wieneke_2015}. The original interrogation window is subdivided into smaller windows for which the averaged particle velocity is evaluated as well. Then, the variance of the coarse-grained velocity and the sub-window velocities can be computed, indicating whether the coarse-grained velocity does approximate the continuum fluid velocity. Physically, this procedure corresponds to the evaluation of $\boldsymbol{\tau}_{kin}$ from Eq. (\ref{eq:KineticStressSuperFluid}) for $W_h(\mathbf{x} - \mathbf{y}(t)) = 1$. If $\boldsymbol{\tau}_{kin} = \mathbf{0}$, the super fluid does perfectly approximate the actual, corresponding fluid elements and the coarse grained interrogation window is sufficient in terms of spatial resolution. \\
    Based on our theory we can not only interpret the sub-window method, we can also ask whether it might be possible to apply the method vice versa? Assuming that the spatial resolution within the sub-windows is sufficient, can we deliberately measure $\boldsymbol{\tau}_{kin}$ in Eq. (\ref{eq:KineticStressSuperFluid}) on the scale of the interrogation window? Although a connection between LES and PIV was already made in the past in order to measure dissipation rates utilizing eddy viscosity models \cite{Sheng_2000}, to our knowledge a thorough PIV analysis of $\boldsymbol{\tau}_{kin}$ and its statistics, based on the Eq. (\ref{eq:KineticStressSuperFluid}), was not conducted yet.  We are convinced that, if possible, it could open new doors in turbulence research.

\end{itemize}

\emph{Conclusion} - We like to complete our work with the conclusion that the LES framework is much more than just a numerical treatment of turbulence. By the introduction of the concept of the super fluid element in combination with Hardy theory, we were able to demonstrate that the transport equations for the super fluid element are equivalent to the governing equations of compressible LES. This result was derived on first principles together with an appropriate averaging process, in which we have interpreted elementary fluid elements as Lagrangian particles. During the mathematical derivation it became evident that Favre filtering is a natural choice in the LES framework and why eddy viscosity models are commonly employed. The latter is justified by our proof demonstrating the equivalence of the kinetic stress tensor for the super fluid element, namely $\boldsymbol{\tau}_{kin}$, and the well-known subgrid stress tensor $\boldsymbol{\tau}_{SGS}$ from LES. This kinetic stress tensor $\boldsymbol{\tau}_{kin}$ could be the foundation of a new quantitative definition of turbulence.


\bibliographystyle{apsrev4-1} 

\end{document}